\documentclass[twocolumn,showpacs,prb,aps]{revtex4}
\usepackage{graphicx,amsmath,amssymb}

\begin{document}
\title{Divergence of the orbital nuclear magnetic relaxation rate in metals}
\author{A.~Knigavko}
\email{anton.knigavko@brocku.ca}
\author{B.~Mitrovi\'c}
\author{K.V.~Samokhin}
\affiliation{Department of Physics, Brock University,
St.~Catharines, Ontario, Canada, L2S 3A1}
\pacs{76.20.+q,76.30.Pk,74.25.Nf}
\date{\today}

\begin{abstract}
We analyze the nuclear magnetic relaxation rate $(1/T_1)_{orb}$ due to
the coupling of nuclear spin to the orbital moment of itinerant electrons
in metals. In the clean non--interacting case, contributions from
large--distance current fluctuations add up to cause a divergence of $(1/T_1)_{orb}$.
When impurity scattering is present, the elastic mean free time $\tau$ cuts off the
divergence, and the magnitude of the effect at low temperatures is controlled by
the parameter $\ln(\mu \tau)$, where $\mu$ is the chemical potential.
The spin--dipolar hyperfine coupling, while has the same spatial variation $1/r^3$
as the orbital hyperfine coupling,
does not produce a divergence in the nuclear magnetic relaxation rate.
\end{abstract}

\maketitle

\section{Introduction}

The nuclear magnetic resonance (NMR) provides powerful experimental
tool in material science. One prominent example in the condensed matter
is that measurements and analysis of the Knight shift and nuclear magnetic
relaxation rate (also called nuclear spin--lattice relaxation rate)
proved to be a decisive test\cite{hebel59} of the validity of the
BCS theory of superconductivity.
In general, the NMR signal carries information on the interactions which
couple the nuclear magnetic moments to their environment. In such a situation
the theoretical understanding of various possible contributions to the total
measured quantity is very important.

For metals the most important hyperfine coupling is the one with the itinerant electrons
\cite{slichter-book,winter-book}.
The magnetic hyperfine interaction of each nucleus with the electrons may be written as
$
    {\cal H}_{hf} = -\gamma_n {\bf I} \cdot {\bf h},
$
where $\gamma_n$ is the nuclear gyromagnetic ratio
and ${\bf I}$ is the spin of the nucleus located at the position $\bf R$.
The units where $\hbar=k_B=1$ are used throughout except in Sec.~VI.
The effective hyperfine magnetic field ${\bf h} $  has three contributions:
\begin{equation}
\label{effective-field-general}
    {\bf h}({\bf R}) = \gamma_e \int {\rm d}^3 r
    \Bigg[
    -\frac{8\pi }{3} {\bf s} \delta (\boldsymbol{\rho})
    + \frac{{\bf s} - 3\hat{\boldsymbol{\rho}}
    ({\bf s}\cdot\hat{\boldsymbol{\rho}})}{\rho^3}
    - \frac{{\bf l}}{\rho^3}
    \Bigg],
\end{equation}
where ${\bf s}$ and ${\bf l}$
are the spin and the orbital moment of the conduction electron located at ${\bf r}$, and
we use the notations $\boldsymbol{\rho}={\bf r}-{\bf R}$, $\rho=|\boldsymbol{\rho}|$
and $\hat{\boldsymbol{\rho}}=\boldsymbol{\rho}/\rho$.
In Eq.~(\ref{effective-field-general}) the integration is over the sample volume
$V$ and  $\gamma_e = e/(m c)$ is the electron
gyromagnetic ratio, $c$ is the speed of light, $-e$ is the electron charge
(with $e>0$) and $m$ is the free electron mass.
The first term in Eq.~(\ref{effective-field-general}) originates from the
Fermi--contact hyperfine interaction, the second term is due to the spin--dipole
hyperfine interaction, and the third term is due to the orbital hyperfine interaction.

In this paper we would like to draw attention to an interesting property
of the nuclear magnetic relaxation rate $(1/T_1)_{orb}$ arising as a result
of orbital hyperfine interaction: For a perfectly clean metallic
system of infinite spatial dimensions, and in the absence of an external magnetic
field, $(1/T_1)_{orb}$ is infinitely large. This fact has been already reported
by Lee and Nagaosa \cite{lee91}. Here we provide a thorough discussion of the
situation using the Green's functions method.
The nature of the effect is related to the properties of long--range
static fluctuations of orbital fields and currents. Mathematically,
the divergence of $(1/T_1)_{orb}$ appears upon integrating the relevant
electronic two--particle correlation function over the momentum transfer {\bf q}:
the integral diverges at ${\bf q}=0$. The divergence is thus related to the
behavior of the electrons that are far away from the nucleus.
In this paper the influence of the finite sample size is included via a
cutoff of the wave vector integrals, and possible surface effects are not
considered explicitly.

The divergence of $(1/T_1)_{orb}$ means that, in principle, in a sufficiently
clean sample the orbital mechanism for the nuclear magnetization relaxation
is very efficient and the total $T_1$ can be very small.
In real material some impurities are always present.
We show that electron scattering off impurities removes the divergence.
The magnitude of $(1/T_1)_{orb}$ at low temperatures is controlled by the
parameter $\ln(\mu \tau)$, where $\mu$ is the chemical potential and
$\tau$ is the mean free time for itinerant electrons.
This means a logarithmic dependence on the impurity concentration.
Our numerical estimates for {\sl Li} and {\sl Sr$_2$RuO$_4$} show that
even for values of $\tau$ characteristic of the cleanest samples
the long--range part of $(1/T_1)_{orb}$ is not dominant. But it can be big
enough, we believe, to be experimentally determined.

The rest of the paper is organized as follows. In Section II we present
the formula for the relaxation rate which is used for the calculation.
In Section III we evaluate $(1/T_1)_{orb}$ in a simple model in which the
normal metal is described by the free electron gas.
We study both clean and impure cases, with the details regarding the vertex
corrections relegated to the Appendix.
In Section IV we show that, unlike the orbital hyperfine interaction, the
long--range contribution to $1/T_1$ due to the spin--dipole hyperfine interaction
is finite.
In Section V we demonstrate that the singularity of $(1/T_1)_{orb}$ in a clean metal
is in fact quite general. First, we show that placing the free electron gas into a
periodic potential (the simplest model for a crystalline solid) does not remove the
singularity.
Next, following the idea of Ref.~\onlinecite{lee91}, we show how metallic
systems with an arbitrary electronic dispersion can be analyzed using
the connection between $(1/T_1)_{orb}$ and the nonuniform static electric
conductivity $\sigma_{\alpha\beta}({\bf q})$. Recognizing the fact that
the static electric conductivity of the perfectly clean electronic system
is infinite perhaps makes the divergence of
$(1/T_1)_{orb}$ in such a system less surprising and puzzling.
Section VI contains a discussion, where
in particular we estimate the magnitude of the effect, and the concluding remarks.

\section{The relaxation rate formula}

The expression for the nuclear magnetic relaxation rate that is most suitable
for our purposes has the following form\cite{moriya56,wolff63,samokhin05}:
\begin{equation}
\label{relax-rate-definition}
\frac{1}{T_1} = -\frac{\gamma_n^2}{2} \coth \left(\frac{\omega_0}{2T}\right)
{\rm Im}  K^R_{+-}(\omega_0, {\bf R}),
\end{equation}
where $\omega_0 = \gamma_n H$ with $H$ being the external magnetic
field oriented along $z$ axis. In the derivation of Eq.~(\ref{relax-rate-definition})
the hyperfine interaction is treated as a perturbation for the
electronic Hamiltonian, which can be quite general.
The main object in Eq.~(\ref{relax-rate-definition}) is the Fourier transform
of the retarded correlator of the effective magnetic fields $\bf h({\bf R})$:
\begin{eqnarray}
\label{ret-field-correlator}
    K^R_{+-}(t,{\bf R}) &=& -i\theta(t)\langle[h_+(t,{\bf R}),h_{-}(0,{\bf R})]\rangle,
\end{eqnarray}
where $h_{\pm} \equiv  h_x \pm i h_y$.
We take the origin of coordinate system at the location of the nucleus
(i.~e. ${\bf R}=0$).

The correlator
$K^R_{+-}(\omega)$ can be computed starting from the explicit
expression for the orbital hyperfine fields in the formalism of
the second quantization. We  follow the standard procedure and find
at first the corresponding Matsubara correlator $K^M_{+-}(i\nu_n)$
and then apply the analytical continuation from the imaginary axis
to just above the real axis:
$K^R_{+-}(\omega) = K^M_{+-}(i\nu_n\rightarrow \omega +i0^{+})$.
Alternatively, using Maxwell equations one can express
the real frequency correlator $K^R_{+-}(\omega)$ through the
current--current correlator and thus through the electrical
conductivity, for which there exist well developed methods of
calculation including the kinetic equation approach and
the linear response theory.

Based on the magnitude of the nuclear magneton, the approximation relevant
to experiments in typical laboratory magnetic fields and not too low
temperatures is $\omega_0 \rightarrow 0$.
Then Eq.~(\ref{relax-rate-definition}) gives
\begin{equation}
\label{relax-rate-w0-definition}
{1 \over T_1 T} =
-\gamma_n^2 \lim_{\omega_0\rightarrow 0} \frac{1}{\omega_0}
{\rm Im}  K^R_{+-}(\omega_0, {\bf R}=0).
\end{equation}
The approximation $\omega_0 \rightarrow 0$ also implies that statistical averages
have to be calculated for the system in zero external magnetic field $H$.

The magnetic field to be inserted in Eq.~(\ref{relax-rate-w0-definition}) is the
effective magnetic field given by Eq.~(\ref{effective-field-general}).
We would like to comment that there is one approximation which is implicit
in presenting the fluctuating magnetic field at nucleus in this form. It
consist in neglecting the influence of the magnetic field on the electrons
themselves. This feedback effect may be phenomenologically accounted for
by replacing ${\bf h}$ on the left hand side of
Eq.~(\ref{H-second-quantization}) with
$(1+4\pi\chi){\bf h}$, where $\chi$ is the magnetic susceptibility
of the material. For ordinary metals $|\chi|\sim 10^{-5}$ and we will not
include this contribution in what follows. The complete treatment of the
feedback effect, which would be necessary when $|\chi|$ is large,
in particular in superconductors,
requires taking into consideration dynamical fluctuations of the
vector potential $\bf A$ of the electromagnetic field. This is beyond the
scope of the present paper.

\section{Calculation of $(1/T_1)_{orb}$ for a simple model of normal metal}

In this subsection we consider in detail the nuclear magnetic relaxation
rate due to orbital hyperfine interaction, $(1/T_1)_{orb}$, in the situation
where electronic system is modeled by the three--dimensional electron gas
(moving in the positively charged uniform background). The electrons do not
interact between themselves, but could elastically scatter
off randomly distributed nonmagnetic impurities.

This model is sufficient to demonstrate that for the relaxation of nuclear magnetization
in the absence of an external magnetic field $(1/T_1)_{orb}$ diverges in the clean case.
This divergence is cut off when impurities are present.
An extension of the argument to more general situations is presented in Section V.

In the
second--quantization representation the operator of the  orbital magnetic
field at the origin ${\bf R}=0$ has the form:
\begin{equation}
\label{H-second-quantization}
\hat{{\bf h}}=
-\gamma_e \int d^3r \sum_\sigma \hat{\psi}^{\dag}_{\sigma}({\bf r})
\frac{ {\bf r} \times (-i {\bf \nabla})}{r^3} \hat{\psi}_{\sigma}({\bf r}),
\end{equation}
which can be obtained from the second term in
Eq.~(\ref{effective-field-general}) by combining the second
quantized representation of the momentum operator $\hat{\bf p}$ with
the definition of the angular momentum. The hats denoting the
operators will be omitted below in order to simplify notation.
Note that expression for the magnetic field is consistent
with the Biot--Savart law.
Indeed, Eq.~(\ref{H-second-quantization}) can be obtained if
in the expression for the magnetic field\cite{jackson-book}
at ${\bf R}=0$, given by the integral
${\bf h}=-(1/c)\int d^3r [{\bf j}({\bf r})\times{\bf r}]/r^3$,
the standard second--quantized
representation for the orbital electric current is inserted
and the integration by parts is performed while neglecting the
surface term.

\subsection{Free electron gas in the plane wave basis}

The $\psi$ operators can be expanded in the basis of plane waves,
which are solutions of the Schr\"{o}dinger equation in empty space:
\begin{equation}
\label{psi-free-electrons-plane waves}
\psi_{\sigma}({\bf r})=\frac{1}{\sqrt{V}} \sum_{\bf k} e^{i {\bf k r}} c_{{\bf k} \sigma},
\end{equation}
where $V$ is the volume of the system. When this expansion
is inserted in the definition of $\bf h$ in Eq.~(\ref{H-second-quantization}) a
straightforward $\bf r$--integration leads to the following expression for
the effective magnetic field due to the electron orbital motion:
\begin{equation}
\label{h-4-free-electrons}
{\bf h} =  \gamma_e \frac{4\pi i}{V} \sum_{{\bf k},{\bf k}^\prime}
\frac{{\bf k} \times {\bf k}'}{({\bf k}-{\bf k}')^2}
\sum_\sigma c^{\dag}_{{\bf k} \sigma} c_{{\bf k}' \sigma}.
\end{equation}
The corresponding Matsubara correlator
$
K^M_{+-}(i\nu_n) = - \int_0^{\beta} d\tau e^{i \tau \nu_n}
\langle T_{\tau} h_+(\tau)h_{-}(0)  \rangle
$
has the form:
\begin{multline}
K_{+-}^M(i\nu_n) = 2 \frac{(4\pi\gamma_e)^2}{V^2}\sum_{{\bf k}_1,{\bf k}_2}
\frac{\left( {\bf k}_1 \times {\bf k}_2 \right)^2_x
+ \left({\bf k}_1 \times {\bf k}_2 \right)^2_y}{|{\bf k}_1-{\bf k}_2|^4}
\\
\times S(i\nu_n,{\bf k}_1,{\bf k}_2).
\label{matsubara-corr-free-el}
\end{multline}
The quantity $S$ is related to the electron bubble:
\begin{multline}
{\cal B}(i\nu_n,{\bf k}_1,{\bf k}'_1,{\bf k}_2,{\bf k}'_2)
\equiv
\int_0^{\beta} d\tau e^{i \tau \nu_n} \sum_{\sigma_1 \sigma_2} \langle
T_\tau c^{\dag}_{{\bf k}_1 \sigma_1}(\tau)
\\
\times
c_{{\bf k}'_1 \sigma_1}(\tau)
c^{\dag}_{{\bf k}_2 \sigma_2}(0) c_{{\bf k}'_2 \sigma_2}(0) \rangle,
\label{cccc-averaging}
\end{multline}
where $\beta=1/T$ with $T$ being temperature. The angular brackets denote
the thermodynamic average with the given electronic Hamiltonian, and
$T_{\tau}$ is the imaginary time ordering operator.
For the free electron gas we have
\begin{equation}\label{free-el-bubble-im-axis-general}
    {\cal B}(i\nu_n,{\bf k}_1,{\bf k}'_1,{\bf k}_2,{\bf k}'_2) =
      -2 \delta_{{\bf k}_1, {\bf k}'_2} \delta_{{\bf k}_2, {\bf k}'_1}
      S(i\nu_n,{\bf k}_1,{\bf k}_2),
\end{equation}
and the quantity $S$ has the following form:
\begin{eqnarray}
    S(i\nu_n,{\bf k}_1,{\bf k}_2) &=&
    T \sum_{m} G_0({\bf k}_1, i\omega_m + i\nu_n) G_0({\bf k}_2, i\omega_m )
    \nonumber \\
    &&= \frac{f(\xi_{{\bf k}_1}) - f(\xi_{{\bf k}_2})}{\xi_{{\bf k}_1} - \xi_{{\bf k}_2} -i\nu_n},
    \label{free-el-bubble-im-axis}
\end{eqnarray}
where $\omega_m=2\pi T(m+1/2)$ and $\nu_m=2\pi T m$, $m=0,\pm 1,...$,
are the fermionic and bosonic Matsubara frequencies, respectively.
Also, in Eq.~(\ref{free-el-bubble-im-axis})
$G_0({\bf k}, i\omega_m )= 1/(i\omega_m - \xi_{\bf k})$
is the free electron Green's function, with $\xi_{\bf k} = \varepsilon_{\bf k}-\mu$,
and $f(x)= 1/[\exp(x/T)+1]$ is the Fermi distribution function.
Note that $S$ depends on the momenta $\bf k$ only through the energies
$\varepsilon_{\bf k}={\bf k}^2/(2m)$.

To arrive at Eq.~(\ref{matsubara-corr-free-el}) the factors with the vector products
of four $\bf k$ vectors in the integrand of the expression for the correlator $K_{+-}^M$
were simplified using the delta functions of Eq.~(\ref{free-el-bubble-im-axis-general}):
\begin{equation}
\label{vect-product-GG}
    \frac{ ({\bf k}_1 \times {\bf k}'_1)_{+} }{ ({\bf k}_1 - {\bf k}'_1)^2}
    \frac{ ({\bf k}_2 \times {\bf k}'_2)_{-} }{ ({\bf k}_2 - {\bf k}'_2)^2} \rightarrow
    -\frac{({\bf k}_1 \times {\bf k}_2)_x^2 + ({\bf k}_1 \times {\bf k}_2)_y^2}
    {|{\bf k}_1 - {\bf k}_2|^4}.
\end{equation}
For the quadratic dispersion of the free electrons the density of electronic states
per one spin projection $N(\varepsilon)= m\sqrt{2m\varepsilon}/(2\pi^2)$
does not depend on angular variables. Converting the $\bf k$--summations into
integrations and separating the angular variables we can write:
\begin{multline}
K_{+-}^M(i\nu_n) =
 2(4\pi \gamma_e)^2  \int_0^{+\infty} d\varepsilon_1 d\varepsilon_2
 N(\varepsilon_1) N(\varepsilon_2)
 \\
 \times K_{\Omega}(\varepsilon_1,\varepsilon_2)
    S(i\nu_n,\varepsilon_1,\varepsilon_2),
    \label{matsubara-corr-free-el-form2}
\end{multline}
where we have introduced the quantity $K_{\Omega}$ defined as an angular integral,
which in the present case contains just one nontrivial integration over
the angle between ${\bf k}_1$ and ${\bf k}_2$ that can be performed
exactly:
\begin{eqnarray}
\label{angular-part-free-electrons}
K_{\Omega}(\varepsilon_1,\varepsilon_2)
 &\equiv&
 \frac{2}{3} \oint \frac{d \Omega_{{\bf k}_1}}{4\pi} \frac{d \Omega_{{\bf k}_2}}{4\pi}
    \frac{\left( {\bf k}_1 \times {\bf k}_2 \right)^2}{|{\bf k}_1-{\bf k}_2|^4}
    \label{angular-part-k-vector} \\
  &=& \frac{1}{6}\left[a \ln \frac{a+1}{a-1} -2 \right].
    \label{angular-part-energy}
    \end{eqnarray}
Here $ k_{1,2}=|{\bf k}_{1,2}|$ and
$a\equiv (k_1^2 + k_2^2)/(2 k_1 k_2)=
(\varepsilon_1 + \varepsilon_2)/(2\sqrt{\varepsilon_1 \varepsilon_2)}$,
which in turn implies  that
$a \pm 1 = (k_1 \pm k_2)^2 /(2 k_1 k_2)
= (\sqrt{\varepsilon_1} \pm \sqrt{\varepsilon_2})^2/(2\sqrt{\varepsilon_1 \varepsilon_2})$.

Now we perform the analytic continuation of $S(i\nu_n)$ given by
Eq.~(\ref{free-el-bubble-im-axis}) on the imaginary axis to the frequencies just above
the real axis and expand in $ \omega \ll \mu$, keeping in mind that we actually need
only the imaginary part for small $\omega$. Introducing the notation
$S(\omega)\equiv S(i\nu_n \rightarrow  \omega +i0^{+})$ we write:
\begin{eqnarray}
{\rm Im} \; S(\omega)
     &\approx& \pi \delta(\xi_1-\xi_2-\omega)\,
    [\omega f'(\xi_2) + {\cal O}(\omega^2)],
    \label{el-bubble-im-part}
\end{eqnarray}
where $f'(\xi)\equiv \partial f(\xi)/\partial\xi$. The delta function in
Eq.~(\ref{el-bubble-im-part}) allows us to eliminate one energy integral in
Eq.~(\ref{matsubara-corr-free-el-form2}), while the derivative of the Fermi
distribution places the energy close to the Fermi surface. This allows us
to use for the degenerate electron gas the standard approximation of the
constant DOS at the Fermi level and switch in Eq.~(\ref{matsubara-corr-free-el-form2})
from the $\varepsilon$ integrations to the $\xi$ integrations from $-\infty$
to $+\infty$. Additionally, the angular factor $K_{\Omega}$ reduces to
\begin{equation}
\label{angular-part-reduced}
    K_{\Omega}(\xi_1,\xi_2) = \frac{1}{3} \ln \frac{\bar{\mu}}{|\xi_1 - \xi_2|},
\end{equation}
where $\bar{\mu}\equiv(4/e)\mu \approx 1.47\mu$.
It is worth emphasizing that the angular integral $K_{\Omega}$ brings in a
nonanalytic energy dependence. Thus, at small $\omega$ we arrive at the following
integral:
\begin{multline}
\lim_{\omega\rightarrow 0}\frac{{\rm Im} K^R_{+-}(\omega)}{\omega}
= \frac{2\pi}{3} [4\pi\gamma_e N(\mu)]^2
    \int_{-\infty}^{+\infty} d\xi_1 d\xi_2 f'(\xi_2)
\\
\times  \delta(\xi_1-\xi_2-\omega)\;
    \ln \frac{\bar{\mu}}{|\xi_1 - \xi_2|}.
\label{ImK-xi-interlas-form}
\end{multline}
The final expression for the relaxation time has the form:
\begin{equation}
\label{rel-rate-free-el-form1}
    {1 \over T_1 T} = \frac{2\pi}{3}
    \left[4\pi \gamma_n \gamma_e N(\mu) \right]^2
     \ln\left( {\bar{\mu} \over \omega} \right),
\end{equation}
where the right--hand side does not depend on temperature.
Note that this expression diverges logarithmically at $\omega \rightarrow 0$, which
can be traced back to the logarithmic behavior of the angular integral $K_\Omega$ in
Eqs.~(\ref{angular-part-energy}) and (\ref{angular-part-k-vector})
at $\varepsilon_1 \rightarrow \varepsilon_2$ or ${\bf k}_1\rightarrow{\bf k}_2$.

\subsection{Free electron gas in the spherical harmonics expansion}

It is instructive to trace the origin of the divergence using the basis of spherical
harmonics around ${\bf R}=0$, where nuclear spin is located. For this purpose we
substitute in Eq.~(\ref{psi-free-electrons-plane waves}) the following
expansion of the plane waves:
\begin{equation}
\label{plane-wave-sp-harmonics}
   \exp(i {\bf k}\cdot{\bf r})=4\pi\sum_{l=0}^{\infty}\sum_{m=-l}^{+l}
    i^l j_l(k r) Y_{lm}^{*}(\hat{\bf k}) Y_{lm}(\hat{\bf r}).
\end{equation}
in terms of the spherical harmonics $Y_{lm}$ and the spherical Bessel functions
$j_l(k r)$ \cite{ryzhik}. An advantage of this approach is that we can use in
Eq.~(\ref{H-second-quantization}) the well known matrix elements of the angular
momentum operator {\bf l}. The radial part can be integrated to give:
\begin{multline}
  h_{\pm} = - \gamma_e  \frac{(4\pi)^2}{V}
  \sum_{{\bf k}_1 {\bf k}_2 \sigma}
  c^{\dag}_{{\bf k}_1 \sigma} c_{{\bf k}_2 \sigma}
  \sum_{l=1}^{\infty} Z_l\left[\frac{4k_1 k_2}{(k_1+k_2)^2}\right]
   \\
  \times \sum_{m=-l}^{+l}\sqrt{(l\pm  m)(l \mp m+1)}
  Y_{l,m \pm 1}^*(\hat{\bf k}_1) Y_{l,m}(\hat{\bf k}_2),
  \nonumber
\end{multline}
\begin{equation}
  Z_l(z)=  \frac{\sqrt{\pi}}{4} \frac{\Gamma(l)}{\Gamma(l+3/2)}
  \left(\frac{z}{4}\right)^l
  F[l,l+1;2(l+1); z],
  \nonumber
\end{equation}
where $\hat{\bf k}={\bf k}/|{\bf k}|$ and $F[a,b;c;z]$ is the hypergeometric function
\cite{ryzhik}.  Calculation of the Matsubara correlator proceeds as in the previous
subsection and produces $S(i\nu_n)$ of Eq.~(\ref{free-el-bubble-im-axis}) and
$S(\omega)$ of Eq.~(\ref{el-bubble-im-part}) after the analytic continuation,
leading to $\delta(\varepsilon_{{\bf k}_1}-\varepsilon_{{\bf k}_2})$ and $k_1=k_2$
at $\omega=0$. At this stage the angular parts of $\bf k$--integrations are performed
and then all the summations except one can be evaluated. The final result reads:
\begin{equation}
\label{rel-rate-free-el-2}
    {1 \over T_1 T} = \frac{2\pi}{3}
    [4\pi\gamma_n \gamma_e N(\mu)]^2
    \sum_{l=1}^{\infty} \frac{16/\pi}{l},
\end{equation}
where finite terms have been omitted. Since we have set up explicitly $\omega=0$
this expression is divergent.  In accordance with what has been found previously in
Eq.~(\ref{rel-rate-free-el-form1}), the divergence is logarithmic and in this case
occurs in the sum over orbital quantum number as $l\rightarrow\infty$.

\subsection{The impurity dependence of $(1/T_1)_{orb}$}

In this subsection we consider the electron gas interacting with
point--like elastic scatterers. The disorder is treated perturbatively
using the standard impurity averaging technique\cite{abrikosov-book}.
In the Appendix we show that in the isotropic situation considered here the
vertex corrections to the impurity averaged electronic bubble, given by
Eq.~(\ref{cccc-averaging}), vanish exactly. Therefore, the impurity
averaged correlator of the orbital magnetic fields $K_{+-}^M(i\nu_n)$
is given by the same expression as in the clean case,
Eq.~(\ref{matsubara-corr-free-el}), with unchanged angular part
$K_{\Omega}$, Eq.~(\ref{angular-part-energy}) but with
\begin{equation}
    S(i\nu_n,{\bf k}_1,{\bf k}_2) =
    T \sum_{m} G({\bf k}_1, i\omega_m
        + i\nu_n) G({\bf k}_2, i\omega_m ),
    \label{el-bubble-im-axis-imp-averaged}
\end{equation}
expressed through the impurity averaged Green's functions:
\begin{equation}
\label{Greens-func-imp}
G({\bf k}, i\omega_m )= \left[i\omega_m - \xi_{\bf k}
        +i\,{\rm sgn}(\omega_m)/(2\tau)\right]^{-1}.
\end{equation}
Here  $1/(2\tau)=\pi N(\mu) n_{imp} V_{imp}^2$ is the elastic scattering
time where $n_{imp}$ is the impurity density and $V_{imp}$ is
the Fourier transform of the impurity potential.
The Matsubara summation in Eq.~(\ref{el-bubble-im-axis-imp-averaged}) can
be conveniently performed using the spectral representation for
the Green's functions. The result is:
\begin{multline}
\lim_{\omega'\rightarrow 0}
\frac{{\rm Im}  S(\omega',\xi_1,\xi_2)}{\omega'}
\\ =
\pi \int_{-\infty}^{+\infty} d\omega  f'(\omega)
A(\xi_1,\omega) A(\xi_2,\omega),
\label{ImS-spectral-presentation}
\end{multline}
where
$
A(\xi,\omega) \equiv A(\xi_{\bf k},\omega)= -{\rm Im}
G({\bf k},i\omega_m\rightarrow\omega+i0^{+})/\pi
$
is the electronic  spectral function in the dirty normal metal
which is sharply peaked around $\xi=\omega$.
The presence of $f'(\omega)$ in Eq.~(\ref{ImS-spectral-presentation}) then
allows us to switch to the $\xi$ integration with the infinite limits in the
expression for the correlator $K^R_{+-}(\omega)$ as in the clean case
(see Eqs.~(\ref{matsubara-corr-free-el-form2}) and (\ref{ImK-xi-interlas-form})).
Now these $\xi$ integrations give:
$
\int_{-\infty}^{+\infty} d\xi_1 d\xi_2 A(\omega,\xi_1) A(\omega,\xi_2)
K_{\Omega}(\xi_1,\xi_2) = \frac{1}{3}\ln(\tau\bar{\mu}),
$
which leads to the following final result:
\begin{equation}
\label{rel-rate-free-el-dirty}
    {1 \over T_1 T} = \frac{2\pi}{3}
     \left[ 4\pi \gamma_n \gamma_e N(\mu) \right]^2
     \ln\left( \tau \bar{\mu} \right).
\end{equation}
We see that $1/\tau$ replaces the frequency $\omega$,
which was present in the correlator for the clean case
(see Eq.~(\ref{rel-rate-free-el-form1})),
and thus removes the divergence at $\omega\rightarrow0$.

\section{Comparison of the orbital contribution to $1/T_1$
with other contributions}

We extend our analysis to include all the hyperfine magnetic fields of
Eq.~(\ref{effective-field-general}), using the same model for electrons
as in the previous section. In the plane--wave basis [see
Eq.~(\ref{psi-free-electrons-plane waves})] one has for three-dimensional
system:
\begin{eqnarray}
  &&{ \bf h}_{F-c} = -\frac{4\pi}{3} \frac{\gamma_e}{V} \sum_{{\bf k},{\bf q}}
  \boldsymbol{\sigma}_{\alpha\beta}\; c^{\dag}_{{\bf k} \alpha} c_{{\bf k}-{\bf q},\beta} ,
  \label{eff-field-q-form-cont}
 \\
  && { \bf h}_{s-d} = 2\pi \frac{\gamma_e}{V} \sum_{{\bf k},{\bf q}}
     \bigg[\frac{{\bf q} ({\bf q}\cdot\boldsymbol{\sigma}_{\alpha\beta}) }{q^2}
     -\frac{\boldsymbol{\sigma}_{\alpha\beta}}{3} \bigg]
     c^{\dag}_{{\bf k} \alpha} c_{{\bf k}-{\bf q},\beta} ,
  \label{eff-field-q-form-dip}
  \\
  && { \bf h}_{orb} = 4\pi \mathrm{i} \frac{\gamma_e}{V} \sum_{{\bf k},{\bf q}}
    \frac{{\bf q} \times {\bf k}}{q^2}
    c^{\dag}_{{\bf k} \alpha} c_{{\bf k}-{\bf q},\alpha} .
 \label{eff-field-q-form-orb}
\end{eqnarray}
In these equations the summations over all repeated spin component indices are assumed.
We observe that the operator of the Fermi--contact effective magnetic field,
${ \bf h}_{F-c}$, and the operator of the spin--dipole effective magnetic field,
${ \bf h}_{s-d}$, can be combined into a single object, for which
the ${\bf q}$--dependence inside the sums appears only through the transverse
projector $\delta_{ij}-q_i q_j/q^2$.
This simplifies the next step of the relaxation rate calculation, which is to
evaluate the Matsubara correlator $K^M_{+-}$. We obtain:
\begin{multline}
K_{+-}^M(i\nu_n) =  2 \frac{(2\pi \gamma_e)^2}{V^2}\sum_{{\bf k},{\bf q}}
\bigg[ 2 - \frac{q_{+}q_{-}}{q^2}
\\
 + 4 \frac{\left( {\bf q} \times {\bf k}\right)_{+}
\left({\bf q} \times {\bf k}\right)_{-}}{q^4} \bigg]
S(i\nu_n,{\bf k},{\bf k}-{\bf q}).
\label{matsubara-corr-normal}
\end{multline}
In the clean case the quantity $S$ is given by Eq.~(\ref{free-el-bubble-im-axis}).
In Eq.~(\ref{matsubara-corr-normal}) the first two terms in the square brackets
represent the combined contribution of the Fermi--contact and the spin--dipole
hyperfine interactions. The third term is the contribution due to the orbital
hyperfine interaction, which was considered before. It is the same as in
Eq.~(\ref{matsubara-corr-free-el})
but written with the change of the summation variables
$({\bf k}_1,{\bf k}_2)\rightarrow({\bf k},{\bf q})$ where
${\bf k} =({\bf k}_1 + {\bf k}_2)/2$, ${\bf q} ={\bf k}_1 - {\bf k}_2$.
In this form the behavior of different contributions as functions of $q=|{\bf q}|$
is clearly displayed. In fact, both Fermi--contact and spin--dipole contributions
depend on $q$ through $S$ function only, while the orbital contribution contains
the additional factor $1/q^2$.

The  divergence in $(1/T_1)_{orb}$ in the clean system appears now as
the divergence of the {\bf q} sum as $q\rightarrow0$. On the other hand,
the spin--dipole hyperfine interaction produces a finite relaxation rate.
This result contradicts expectations\cite{lee91}  that  spin--dipole
and orbital hyperfine interactions should bring about similar contributions
to $1/T_1$ because in the real space they both vary with the distance from
the nucleus as $1/r^3$ [see Eq.~(\ref{effective-field-general})].

\section{General character of the divergence of $(1/T_1)_{orb}$ in clean metals}

In this section we demonstrate that the divergence of $1/T_1$ in a clean metal
is quite general and is not an artifact of the free electron model
we used in previous sections. We now concentrate on the orbital hyperfine
interaction that is responsible for the effect.
First we consider the electron gas in an arbitrary
periodic potential, and show that the divergence does not disappear. Then we relate
the correlator of orbital magnetic fields to the current--current correlator, and
therefore to the electrical conductivity. This formulation allows one to treat an
arbitrary electronic dispersion quite generally. The divergence is shown to be
a consequence of the behavior of the static nonuniform conductivity in the long
wavelength limit.

\subsection{The electron gas in a periodic potential}

Now add to the Hamiltonian of the clean Fermi gas a periodic potential,
and consider the effect it has on the correlator of orbital magnetic fields in
the definition of $1/T_1$ in Eq.~(\ref{relax-rate-definition}).
The solutions to the corresponding Schr\"{o}dinger equation are the Bloch functions which
can be used as the basis for the second--quantized description:
\begin{equation}
\label{psi-free-electrons-Bloch-func}
     \psi_{\sigma}({\bf r})=\frac{1}{\sqrt{V}} \sum_n \sum_{\bf k}
     e^{i {\bf k r}} u_{n{\bf k}}({\bf r})c_{n{\bf k}\sigma},
\end{equation}
where $n$ enumerates the electron bands and the wave vector {\bf k}
is now limited to the first Brillouin zone.
Functions $u_{n{\bf k}}({\bf r})$ are periodic in {\bf r} and can be
expanded in the Fourier series:
\begin{equation}
\label{u-expansion}
    u_{n{\bf k}}({\bf r})=\sum_{{\bf G}}U_{n{\bf G}}({\bf k})e^{i{\bf G}\cdot{\bf r}},
\end{equation}
where {\bf G} is a reciprocal lattice vector. The operator of the effective magnetic
field at the origin, ${\bf R}=0$, due to the electron orbital motion has the form:
\begin{eqnarray}
{\bf h} &=&  \gamma_e \frac{4\pi i}{V}
\sum_{{\bf k}{\bf k}^\prime} \sum_{{\bf G}{\bf G}^\prime}
\frac{({\bf k}+{\bf G}) \times ({\bf k}'+{\bf G}')}
{({\bf k}+{\bf G}-{\bf k}'-{\bf G}')^2}
\nonumber \\
&& \times \sum_{n n'} U_{n{\bf G}}({\bf k})^{*} U_{n'{\bf G}'}({\bf k}')
\sum_{\sigma} c^{\dag}_{n{\bf k} \sigma} c_{n'{\bf k}' \sigma}.
\label{h-4-lattice-electrons}
\end{eqnarray}
This expression is to be compared with Eq.~(\ref{h-4-free-electrons}). The Green's functions
of the electrons in a periodic potential are diagonal in {\bf k} as well as in the spin and
band indices. Using these properties we calculate the correlator of the orbital
magnetic fields at the location of the nucleus.
The expression for the relaxation time  in the limit $\omega\rightarrow 0$
can be written as follows:
\begin{eqnarray}
 {1 \over T_1 T} &=& 2\pi \frac{\big(4\pi\gamma_n\gamma_e\big)^2}{V^2}
\sum_{{\bf k}_1{\bf k}_2} \sum_{n_1 n_2}
\delta(\varepsilon_{n_1,\bf k_1}-\varepsilon_{n_2,\bf k_2})
\nonumber \\
&& \times f'(\varepsilon_{n_2,\bf k_2})
\sum_{\{\bf G\}} {\cal U}^4 {\bf K}_{1+} {\bf K}_{2-},
\label{1overT-lattice-electrons}
\end{eqnarray}
where $\{{\bf G} \}= \{{\bf G}_1,{\bf G}_2,{\bf G}_3,{\bf G}_4\}$ and
\begin{eqnarray}
{\cal U}^4 &=&
U^*_{n1,{\bf G}_1+{\bf G}_2}({\bf k}_1) U_{n1,{\bf G}_4}({\bf k}_1)
\nonumber \\
&&\times U^*_{n2,{\bf G}_3+{\bf G}_4}({\bf k}_2) U_{n2,{\bf G}_2}({\bf k}_2),
\label{U4factor}
\\
{\bf K}_1 &=&  \frac{({\bf k}_1-{\bf k}_2+{\bf G}_1)\times({\bf k}_2+{\bf G}_2)}
{({\bf k}_1-{\bf k}_2+{\bf G}_1)^2},
\\
{\bf K}_2 &=& - \frac{({\bf k}_2-{\bf k}_1+{\bf G}_3)\times({\bf k}_1+{\bf G}_4)}
{({\bf k}_2-{\bf k}_1+{\bf G}_3)^2}.
\end{eqnarray}

Eq.~(\ref{1overT-lattice-electrons}) is an exact expression. We show that
it contains an infinite term.
To separate the divergent contribution we restrict the sums  in
Eq.~(\ref{1overT-lattice-electrons}) by the conditions
${\bf G}_1={\bf G}_3=0$ and $n_1=n_2$, and change the summation variables
${\bf k}_1 ={\bf k} + {\bf q}/2$, ${\bf k}_2 ={\bf k} - {\bf q}/2$.
Then we perform the small ${\bf q}$-- expansion of the arguments of
the delta function and $f'$ in Eq.~(\ref{1overT-lattice-electrons}),
as well as of the ${\cal U}^4$ of Eq.~(\ref{U4factor}). Accounting only
for the leading term in these expansions Eq.~(\ref{1overT-lattice-electrons})
is reduced to:
\begin{eqnarray}
{1 \over T_1 T} &=& 2\pi \frac{\big(4\pi\gamma_n\gamma_e\big)^2}{V^2}
\sum_{{\bf k},{\bf q}}\sum_n \delta({\bf q}\cdot{\bf v}_n({\bf k}))
f'(\varepsilon_n({\bf k}))
\nonumber \\
&& \times \frac{m^2}{q^4}
\left[{\bf q}\times {\bf v}_n({\bf k}) \right]_{+}
\left[-{\bf q}\times {\bf v}_n({\bf k})\right]_{-},
\label{lattice-electron-q-divergence}
\end{eqnarray}
where we have used the following expression\cite{kittel-book} for
velocity of the band electrons:
\begin{equation}
{\bf v}_n({\bf k})
= \frac{1}{m}\sum_{\bf G} ({\bf k}+{\bf G})|U_{n{\bf G}}({\bf k})|^2
= \frac{\partial\varepsilon_n({\bf k})}{\partial{\bf k}}.
\end{equation}
To estimate the {\bf q}-- sum in Eq.~(\ref{lattice-electron-q-divergence})
we change it, for each band $n$, to the integral over the sphere of
an appropriate volume. The integration over $q=|{\bf q}|$ extends from
$q_{min}=0$ to $q_{max}\simeq 2 q_{F,n}$, where $q_{F,n}$ is
of the order of the average separation between two points on the Fermi
surface in the $n$ band. The final result has the form:
\begin{multline}
  {1 \over T_1 T} = \Big(\frac{e\gamma_n}{c}\Big)^2
  \frac{ 4\pi}{V}  \sum_{{\bf k},n}  D_n f'(\varepsilon_n({\bf k}))
\\ \times
  \frac{[{\bf v}_n({\bf k})]^2+[\hat{z}\cdot{\bf v}_n({\bf k})]^2}
  {|{\bf v}_n({\bf k})|},
\label{1overT1-velocity-average-form}
\end{multline}
where $D_n=\ln(2q_{F,n}/q_{min})$ is the divergent factor.
As long as the electrons do not interact with each other $q_{min}=0$
at any temperature. The reason is that the delta function, which appears
in Eqs.~(\ref{1overT-lattice-electrons})
and (\ref{lattice-electron-q-divergence}) and ultimately brings in the
divergence, is in fact the electronic spectral function of the noninteracting
system.
Thus, the same logarithmic divergence as found in Sec. III for free electrons
appears here. For finite samples of the linear size $L$ we have $q_{min}=1/L$.
When impurities are present, from the result of Sec. III~C it is expected
that $q_{min}\simeq 1/(v_F \tau)$, which is in general different for different bands.
In the next subsection we confirm this form of $q_{min}$ for one band with an arbitrary
electronic dispersion.

To conclude this subsection we note that for band electrons we found the
logarithmically divergent contribution to $(1/T_1)_{orb}$ and the factor
in front of it (see Eq.~(\ref{1overT-lattice-electrons})), which
depends only on the averages of the Fermi velocities and does not involve
the position of the nuclear spin in the unit cell.

\subsection{Expressing $K^R_{+-}(\omega,{\bf q})$ in terms of
the current--current correlator}

We now use another method to demonstrate the singular behavior
of the nuclear magnetic relaxation rate due to orbital hyperfine interaction.
It also allows us to obtain a useful expression for $(1/T_1)_{orb}$ for
an anisotropic metal.
The idea, first used by Lee and Nagaosa~\cite{lee91}, is to express the
correlator of effective orbital magnetic fields
$K^R_{+-}(\omega,{\bf R}=0)= \sum_{\bf q} K^R_{+-}(\omega,{\bf q})$,
which appears in Eq.~(\ref{relax-rate-w0-definition}),
through the current--current correlator,
and consequently through the nonuniform electrical dc conductivity.

The fluctuating magnetic fields are calculated using the Maxwell
equation, assuming that the electric currents are given:
$
{\bf \nabla} \times {\bf h} (\omega,{\bf r})=
\bigl(4\pi {\bf j}(\omega,{\bf r}) - i\omega {\bf E}(\omega,{\bf r})\bigr)/c.
$
In metals it is a good approximation to neglect the second term
containing the electric field ${\bf E}$.
Then, applying the Fourier transform to find the magnetic field we get
$
{\bf h}(\omega,{\bf q})= (4\pi i/c)[{\bf q}\times{\bf j}(\omega,{\bf q})]/q^2,
$
and the required retarded correlator of the orbital magnetic fields is
given by the following expression:
\begin{multline}
  K^R_{+-}(\omega,{\bf q}) =  \Bigl( \frac{4\pi}{c} \Bigr) ^2
  {q_m q_k \over q^4}  \bigl[ \epsilon_{xmi}\epsilon_{xkj}
   + \epsilon_{ymi}\epsilon_{ykj}
    \\
   - i \epsilon_{zlp}\epsilon_{lmi}\epsilon_{pkj} \bigr]
   \Pi_{ij}(\omega,{\bf q}),
  \label{field-correlator-via-cur-correlator}
\end{multline}
where $\Pi_{ij}(\omega,{\bf q})$ is the retarded current-current correlator,
$\epsilon_{ijk}$ is the totally antisymmetric tensor
and summation over the repeated indices, which run through $x,y$ and $z$,
is assumed. In the end we are interested in the situation when the constant
external magnetic field is absent. Therefore we take the $\Pi_{ij}$ tensor to be
symmetric, namely $\Pi_{ij}=\Pi_{ji}$. In this case the third term in
the square brackets in Eq.~(\ref{field-correlator-via-cur-correlator}) vanishes.

In general $\Pi_{ij}$ contains both paramagnetic and
diamagnetic terms. However, for the calculation of $1/T_1$ we need
only ${\rm Im} \Pi_{ij}$ (see Eq.~(\ref{relax-rate-definition})),
and the diamagnetic part, which is real, drops out.
The  retarded current-current correlator can be related to the electrical
conductivity tensor\cite{abrikosov-book}. Using
$
{\rm Im} \Pi_{ij}(\omega,{\bf q}) = -{\rm Re} [\omega \sigma_{ij}(\omega,{\bf q})]
$
we obtain from Eq.~(\ref{field-correlator-via-cur-correlator}) the following
expression:
\begin{multline}
  \lim_{\omega\rightarrow 0}  \frac{ {\rm Im} K^R_{+-}(\omega,{\bf q})}{\omega} =
   -\Bigl(  {4\pi \over c}\Bigr)^2 {1 \over q^4}
   {\rm Re}  \big[ (q_x^2+q_y^2)\sigma_{zz}
   \\
  + q_z^2 (\sigma_{xx} + \sigma_{yy} )
   -2q_x q_z \sigma_{xz} -2q_y q_z \sigma_{yz}\big],
  \label{K-no-mag-field}
\end{multline}
where arguments of the conductivity tensor $\sigma_{ij}(\omega=0,{\bf q})$
are not shown explicitly. The expression~(\ref{K-no-mag-field}) has to be inserted
in Eq.~(\ref{relax-rate-w0-definition}).

In the simple case of an isotropic system the conductivity tensor
is diagonal in the coordinate frame with the $z$ axis parallel to
${\bf q}=(q,\theta,\phi)$, namely
${\rm Re}\,\sigma_{ij}(q)={\rm diag}
[\sigma_{\perp}(q),\sigma_{\perp}(q),\sigma_{||}(q)]$.
This tensor should be transformed to the coordinate
frame defined by the initial orientation of the nuclear magnetic moment
(i.e $\hat{z}||{\bf H}$) using an appropriate rotation of coordinate
frame and then inserted into Eq.~(\ref{K-no-mag-field})
and Eq.~(\ref{relax-rate-w0-definition}).
The final expression for nuclear magnetic relaxation rate contains only
the perpendicular conductivity $\sigma_\perp(q)$:
\begin{equation}
\label{relax-rate-form1}
    {1\over T_1 T}
    =\frac{8}{3}  \biggr( \frac{\gamma_n}{c}\biggr)^2
    \int_0^\infty d q \sigma_\perp(q).
\end{equation}
It is known from the theory of the anomalous skin effect \cite{kittel-book}
that in a clean metal
the nonuniform conductivity $ \sigma_\perp(q)$ behaves as $1/q$ at small wave
vectors $q$. This leads to the logarithmic divergence for the nuclear magnetic
relaxation rate $(1/T_1)_{orb}$.

In order to clarify this statement and obtain a useful formula for $(1/T_1)_{orb}$
for electrons with an arbitrary dispersion in the presence of impurities,
the $q$--dependent electric conductivity tensor is now evaluated using
the kinetic equation approach. Electric current is given by
\begin{equation}
\label{current-4-kin-eq}
{\bf j}(\omega,{\bf q}) = 2e \sum_{\bf k} {\bf v_k} n(\omega,{\bf k,q}),
\end{equation}
where ${\bf v}\equiv {\bf v_k}=\partial \xi_{\bf k} /\partial {\bf k} $
is the electron group velocity, and the factor 2 is due to the spin.
The distribution function $n(\omega,{\bf k,q})=f(\xi_{\bf k})+\delta n(\omega,{\bf k,q})$ is
a sum of the equilibrium Fermi distribution  $f$ and a correction $\delta n$ proportional to
the small perturbing electric field ${\bf E(q)}={\bf E}$.
Within the relaxation time approximation for the impurity collision integral we have:
\begin{equation}\label{kin-eq-general}
    \left[ 1/\tau -i(\omega - {\bf q}\cdot{\bf v}) \right]\delta n
        = - e (\partial f/\partial\xi) {\bf v}\cdot{\bf E},
\end{equation}
leading to
\begin{equation}
\label{distrib-func}
\delta n(\omega,{\bf k,q}) =
- \frac{e \tau (\partial f/\partial\xi) {\bf v} \cdot {\bf E}}
{ 1- i\tau (\omega -{\bf q}\cdot{\bf v})}.
\end{equation}
The final result for the conductivity reads:
\begin{eqnarray}
\sigma_{ij}(\omega,{\bf q}) &=& 2e^2 \int_{-\mu}^{\infty} d\xi
\left(-\frac{\partial f}{\partial\xi}\right)
\oint \frac{d \Omega_{\bf k}}{4\pi}  N(\xi,\Omega_{\bf k})
\nonumber \\
&& \times {\tau \left( 1 +i\tau\omega \right)v^i v^j
\over 1+\tau^2 \left( \omega -{\bf q}\cdot{\bf v} \right)^2}.
\label{cond-4-kin-eq}
\end{eqnarray}
This form of $\sigma_{ij}(\omega,{\bf q})$ is valid for an arbitrary
electronic band $\xi_{\bf k}$, which enters through the angle dependent
density of electronic states $N(\xi,\Omega_{\bf k})$.
Using the form of the conductivity tensor form Eq.~(\ref{cond-4-kin-eq})
in Eq.~(\ref{K-no-mag-field}) we arrive at the following expression for
$1/T_1$ due to the orbital hyperfine interaction:
\begin{eqnarray}
 \frac{1}{T_1 T} &=&
 \frac{8}{\tau} \left(\frac{e \gamma_n}{c}\right)^2
  \int_0^{\infty} \frac{d q}{2\pi^2}
 \int_{-\mu}^{\infty} d\xi \left(-\frac{\partial f}{\partial\xi}\right)
 \nonumber \\
 && \times \left\langle N(\xi,\Omega_{\bf k})
 \frac{v_z^2 + v^2 \hat{q}_z^2 -2(\hat{q}_z v_z)(\hat{\bf q}\cdot{\bf v})}
    {1/\tau^2+  q^2 (\hat{\bf q}\cdot{\bf v})^2}
    \right\rangle_{\Omega_{\bf k},\Omega_{\bf q}},
    \label{relax-rate-form-general}
\end{eqnarray}
where the $z$ direction is set by the initial orientation of the
nuclear polarization. The angular brackets denote averaging over
both $\bf k$-- and $\bf q$-- angular variables.
The magnitude of the group velocity $\bf v\equiv v_k $ is fixed at
the Fermi velocity by the derivative of the Fermi distribution.
For an anisotropic system the Fermi velocity is angle dependent.
For electrons with quadratic dispersion the calculation
of $1/T_1$ based on Eq.~(\ref{relax-rate-form-general})
gives the result identical to that given by Eq.~(\ref{relax-rate-form1}).

The $q$ integral should be cut off by  $2q_F$ at the upper limit, which
is necessary in the present approach due to the fact that the kinetic
equation does not capture the large--$q$ behavior of the nonuniform
conductivity correctly. On the other hand, the $q$
integral is convergent at the lower limit. The clean limit corresponds to
$1/\tau \rightarrow 0$. In this case the right hand side of
Eq.~(\ref{relax-rate-form-general}) contains the following $\delta$ function:
\begin{equation}
\label{clean-limit}
    \frac{1/\tau}{1/\tau^2 + ({\bf q}\cdot{\bf v})^2} \rightarrow
    \pi\delta[q(\hat{\bf q}\cdot{\bf v})] =
    \frac{\pi}{q}\delta[\hat{\bf q}\cdot{\bf v}].
\end{equation}
After the integration over $\Omega_{\bf q}$ is performed we can see explicitly
that the remaining $q$ integral diverges as $1/q$ at the lower limit.

\section{Discussion}

It was found experimentally\cite{vandeeeerklink00} that the nuclear
magnetic relaxation in many metals is dominated by the the Fermi--contact
hyperfine interaction, for which
$(1/T_1T)_{F-c}=(4\pi k_B/9) \big[4\pi\gamma_n\gamma_e \hbar^2 N(\mu)\big]^2
  {\langle|u_{\bf k}({\bf 0})|^2\rangle^2_{FS}}$
[in this Section we restore  $k_B$ and $\hbar$].
The situations where this is not so and the
contributions of hyperfine orbital and/or spin--dipole interactions are
significant are  worth of detailed analysis. Loosely speaking, the physical
reason for the large magnitude of the Fermi--contact interaction
is that the conduction band usually contains a large portion
of the atomic $s$--orbitals which give a large contribution to the
electronic spin density at the  nuclear site, specified by the overlap factor
$\langle|u_{\bf k}({\bf 0})|^2\rangle^2_{FS}$. The relative
importance of the other hyperfine interactions increases when the density
of states at the Fermi level  is dominated by the $d$-- and $f$-- bands
as in transition  metals and their
compounds\cite{obata63,clogston64,narath67,asada81},
or when the conduction band happens to be almost free
of $s$ orbitals \cite{pavarini01,antropov93,pavarini06}.

We would like to emphasize that both orbital and spin--dipole hyperfine
contributions to  $1/T_1$ commonly discussed in the literature are
in fact the ``local'' contributions which we did not touch upon in this paper.
It is usually argued that since the orbital and spin--dipole  hyperfine
interactions decay fast enough with distance
[they both vary as $1/r^3$, see Eq.~(\ref{effective-field-general})],
the nuclear spin of a given atom is only affected by the magnetic fields
generated by the electron orbitals centered at this very atom.
In contrast, the contribution we analyzed in the previous sections comes
from large distances. It has to be added to the local orbital contribution.

Below we attempt to estimate the magnitude of the relaxation rate due to the
long--range part of the orbital hyperfine interaction
[denoted $(1/T_1)_{orb}^{l-r}$] in some metals, and compare it with the
contributions due to local part of the orbital hyperfine interaction
[denoted $(1/T_1)_{orb}^{loc}$] and Fermi--contact interaction.
We first consider alkaline metals. In this case, the estimates based on the
quadratic dispersion produce the correct order of magnitude for $(1/T_1)_{F-c}$,
as compared to experimental values. This makes our further estimates of
$(1/T_1)_{orb}^{l-r}$ trustworthy.
Next we discuss Sr$_2$RuO$_4$, the compound which is likely to be suitable
to study the orbital nuclear magnetic relaxation due to the peculiarities
of its electronic structure. Additionally, in this case very clean samples
with a long mean free path are available.
The results of our estimates are summarized in Table~\ref{table1}.

\begin{table}
  \centering
  \begin{tabular}{|c|c|c|c|c|c|}
    \hline
    material & $\frac{1}{T_1T}\Big|_{F-c}$
    & $\frac{1}{T_1T}\Big|_{orb}^{l-r}$
   & $\frac{1}{T_1T}\Big|_{orb}^{loc}$
   & $\frac{(1/T_1)_{orb}^{l-r}}{(1/T_1)_{F-c}}$
   & $\frac{(1/T_1)_{orb}^{l-r}}{(1/T_1)_{orb}^{loc}}$
   \\ \hline
    {\sl Li} & 0.019 & 4.05$\cdot$10$^{-4}$ & 1.14$\cdot$10$^{-3}$
    & 2.2$\cdot$10$^{-2}$ & 0.37 \\
    {\sl Na} & 0.213 & 1.20$\cdot$10$^{-4}$ & 0 & 5.6$\cdot$10$^{-4}$ & -- \\
    {\sl Sr$_2$RuO$_4$} & 1.1$\cdot$10$^{-4}$ & 1.3$\cdot$10$^{-5}$
    & 8.5$\cdot$10$^{-4}$ & 0.12 & 0.015
   \\ \hline
  \end{tabular}
  \caption{
  Numerical values of various contribution to $1/(T_1T)$
  [in the units $(K\cdot sec)^{-1}$] for some metals obtained
  as described in the text. For {\sl Sr$_2$RuO$_4$} the data are
  related to $^{17}$O   in the position O2, which is out of the
  conducting basal plane.}
  \label{table1}
\end{table}

\subsection{Alkaline metals}

We consider {\sl Na} and {\sl Li} metals. For both of them  the Fermi surface
lies within the first Brillouin zone and the electronic dispersion can be taken
$\varepsilon({\bf k})=\hbar^2{\bf k}^2/(2m^*)$ for all energies up to the Fermi
level\cite{footnote2}. This means that the results of Section III~C are directly
applicable.
Numerical values are  $m^*/m=1.45$, $q_F=1.3\cdot10^8 $cm$^{-1}$ for {\sl Li},
and $m^*/m=0.98$, $q_F=9.2\cdot10^7 $cm$^{-1}$ for {\sl Na}, as quoted in
Ref.~\onlinecite{pines55}.
The nuclear magnetic moments are $3.26\mu_N$ for $^7$Li and
$2.22\mu_N$ for $^{23}$Na, where $\mu_N$ is the nuclear magneton.

In the case of the Fermi--contact interaction  the necessary overlap factors
were obtained using the results of the first principle calculations of
Kohn\cite{kohn54} and Kjeldaas and Konh\cite{kjeldaas56}:
16$^2$ for {\sl Li} and 144$^2$ for {\sl Na}.
Thus, our theoretical estimate for {\sl Li} is
$(1/T_1T)_{F-c}\approx 0.019 (K\cdot sec)^{-1}$, which is to be compared to
experimental values $0.023 (K\cdot sec)^{-1}$ of Ref.~\onlinecite{anderson59}
or $0.06 (K\cdot sec)^{-1}$ of more recent work in Ref.~\onlinecite{juntunen05}.
In the case of {\sl Na} our theoretical estimate is
$(1/T_1T)_{F-c}\approx 0.213 (K\cdot sec)^{-1}$, while the experimental
value from Ref.~\onlinecite{anderson59} is $0.196 (K\cdot sec)^{-1}$.

The expression for the nuclear magnetic relaxation due to the long--range part
of orbital hyperfine interaction for quadratic dispersion can be obtained
from Eq.~(\ref{rel-rate-free-el-dirty}) and
Eq.~(\ref{1overT1-velocity-average-form}) as follows:
$(1/T_1T)_{orb}^{l-r}=(2\pi/3) \big[4\pi\gamma_n\gamma_e \hbar^2 N(\mu)\big]^2
(m/m^*)^2  \ln(2 q_F l)$.
For our estimates we use $l=10^{-5}$cm and obtain that
$(1/T_1T)_{orb}^{l-r}=4.05\cdot10^{-4}(K\cdot sec)^{-1}$ in {\sl Li} and
$1.20\cdot10^{-4}(K\cdot sec)^{-1}$ in {\sl Na}.
We present here a useful formula that defines the ratio of the two discussed
relaxation rates:
\begin{eqnarray}
\label{estimate}
    \frac{(1/T_1)_{orb}^{l-r}}{(1/T_1)_{F-c}} &= &
      \frac{3}{2} \left(\frac{m}{m^*}\right)^2
      \frac{\ln\big(2q_F l\big)}{\langle|u_{\bf k}({\bf 0})|^2\rangle^2_{FS}}.
\end{eqnarray}
For {\sl Li} this ratio is 2.2\% while for {\sl Na} it is 0.056\%.
Such large difference is explained by the fact\cite{kohn54} that the wave
function of {\sl Li} at the Fermi level contains about 73\% of the $l=1$
spherical harmonic, and only 20\% of the spherically symmetric $l=0$ harmonic.

The magnitude of
the local part of the orbital magnetic nuclear relaxation in both metals
was obtained experimentally by Hecht and Redfield \cite{hecht63} from the
measurements of the Overhauser effect. They found that
$\big[(1/T_1)_{orb}^{loc}\big]/\big[(1/T_1)_{F-c}\big]\approx0.06\pm0.03$
in {\sl Li}, while there is no local contribution in {\sl Na}.
Thus, the local and long--range contributions to $1/(T_1)_{orb}$ appear
comparable in {\sl Li}, and both are quite noticeable.

It appears that metallic {\sl Li} could be the system of choice to look
for the long--range orbital contribution to the nuclear magnetic relaxation.
One can use the characteristic logarithmic dependence of $(1/T_1T)_{orb}^{l-r}$
on the impurity concentration.

It is worth noting that  the NMR experiments are often done with collections
of relatively small metallic particles\cite{anderson59}.
This is dictated by the experimental requirement that a large volume of the
sample should be subjected to a radio frequency magnetic field which only
penetrates within the skin depth. Such experimental setup would limit the
magnitude of the long--range orbital effect even for the very clean metals.

\subsection{\sl Sr$_2$RuO$_4$}
The electronic properties of this material are well studied\cite{mackenzie03},
both theoretically and experimentally, in connection with unconventional
superconductivity, which appears at temperature of about 1K. For the purpose of
a rough estimate for the long--range orbital magnetic relaxation rate we use
the model of three two dimensional bands, with cylindrical Fermi surfaces and
$q_{F,1}=3.04\cdot10^{7}$cm$^{-1}$, $q_{F,2}=6.22\cdot10^{7}$cm$^{-1}$ and
$q_{F,3}=7.53\cdot10^{7}$cm$^{-1}$, as quoted in Ref.~\onlinecite{mackenzie03}.
The lattice parameter along the tetragonal symmetry
axis is $a_3=1.27\cdot10^{-7}$cm. To obtain a convenient formula from
Eq.~(\ref{1overT1-velocity-average-form}) we assume that $v_z^2/(v_x^2+v_y^2)\ll 1$
and the cutoff wave vector $q_{min}$ is the same  for all bands.
Using the density of the electronic states for each band in the form
$N_{2D}(\mu)=m^*/(2\pi\hbar^2 a_3)$ and taking $\hat{z}$ along the tetragonal
symmetry axis we obtain:
\begin{equation}
\label{1overT-2D}
    \left.\frac{1}{T_1T}\right|_{orb}^{r-l}=
    \frac{2k_B}{\hbar}\Big(\frac{e\gamma_n}{c}\Big)^2
    \sum_{n=1}^3 \frac{q_{F,n}}{a_3} \ln(2q_{F,n} l).
\end{equation}
The best samples of {\sl Sr$_2$RuO$_4$} used in the de Haas -- van Alphen
experiments\cite{mackenzie03}  are very clean, with the elastic mean
free path reaching $3\cdot10^{-5}$cm. For the relaxation of the magnetic moment
of $^{17}O$ nucleus, with $\gamma_n=-3.63\cdot10^{3}(sec\cdot G)^{-1}$, we
obtain $(1/T_1T)_{orb}^{r-l}=1.3\cdot10^{-5}(K\cdot sec)^{-1}$. This
value can be compared with the local part of the orbital
nuclear relaxation rate. For example, from the band structure calculations
Pavarini and Mazin \cite{pavarini06} found that
$(1/T_1T)_{orb}^{loc}=8.5\cdot10^{-4}(K\cdot sec)^{-1}$ for the so called
O2 oxygen, located out of the basal conducting planes. The ratio is
$\big[(1/T_1)_{orb}^{l-r}\big]/\big[(1/T_1)_{orb}^{loc}\big]=0.015$.

The O2 position of the oxygen is very favorable for studies of the
long range orbital mechanism of the nuclear magnetic relaxation, because
the local orbital contribution is much smaller than for the other oxygen site, O1.
The Fermi contact contribution is also numerically quite small for the O2 position,
$(1/T_1T)_{F-c}=1.1\cdot10^{-4}(K\cdot sec)^{-1}$
as calculated in Ref.~\onlinecite{pavarini06}. It is still larger than the
long--range orbital contribution, but not substantially:
$\big[(1/T_1)_{orb}^{l-r}\big]/\big[(1/T_1)_{F-c}\big]=0.12$.

Comparison of our predictions with experiments is complicated by the fact that
the experimental values of $1/T_1T$ for O2 oxygen\cite{mukuda98} are two orders
of magnitude greater than values given by the band structure calculations.
The experimental results  necessarily include the effect of electronic correlations,
which are strong in {\sl Sr$_2$RuO$_4$}. Such effects were completely ignored in our
discussion of $(1/T_1)_{orb}^{r-l}$ in this paper.

\subsection{Concluding remarks}
We would like to point out that besides the elastic impurity scattering, other
scattering mechanisms for electrons should also affect to $(1/T_1)_{orb}$.
One expects that inelastic scattering, for example by phonons,
should lead to $1/(T_1 T)\sim \ln T$, because of  the temperature dependence
of the inelastic scattering rate.

The NMR relaxation mechanism through the orbital hyperfine interaction that
we have discussed in this paper is quite general. It is not limited to metals
and should exist in any system with mobile charge carriers. On the basis of
Eq.~(\ref{estimate}) one expects that a small effective mass of the
charge carriers should significantly enhance the effect. Small masses are
typically found in semiconductors \cite{hewes73}. In such systems the NMR
techniques have also been very successful, though the analysis should be
modified compared to the case of metals \cite{selbach79}.

Finally, we would like to comment on calculations of $(1/T_1)_{orb}$ from the
first principles using band structure methods. These calculations typically use
the expansions in a set of basis functions that involve the spherical harmonics
centered at the nuclear site, and an appropriate radial part obtained from
the solution of the Schr\"{o}dinger equation inside the Wigner--Seitz sphere
or the muffin-tin sphere\cite{narath67,asada81,antropov93}.
The series in the orbital quantum number $l$ is truncated at some stage of the
computation based on an empirical estimate of convergence. However, we have
demonstrated in this paper that this expansion is expected to diverge as
$\ln l$, which is usually difficult to capture numerically.
The presence of this divergence calls for a careful examination
of the conditions used to justify the truncation of the $l$ expansion.

In conclusion, in this paper we have discussed the nuclear magnetization
relaxation rate due to the orbital hyperfine interaction of nuclear spins with
itinerant electrons. For an infinite clean metal $(1/T_1)_{orb}$ would become
divergent at zero temperature. The reason for the divergence is that the total
fluctuating magnetic field created by all the electrons at the site of a given
nucleus contains a contribution from distant electrons.
The divergence is removed when there are scattering mechanisms in the system.
At low temperatures the scattering is predominantly elastic.
In this case $(1/T_1)_{orb}$ depends logarithmically on impurity concentration.
Based on our estimates, it seems to be feasible to observe this effect
e.~g. in metallic {\sl Li}.

\section{Acknowledgements}

This work is supported by the Natural Science and Engineering Research
Council of Canada (NSERC).

\appendix
\section{Impurity averaged correlator of orbital magnetic fields}

The operator of the magnetic field created by the orbital motion of the
electrons at ${\bf R}=0$ is given in Eq.~(\ref{h-4-free-electrons}). Here we want
to compute the impurity averaged correlator of such orbital magnetic fields
at a given Matsubara frequency $i\nu_n$. This quantity can be written as follows:
\begin{multline}
  \langle K_{+-}^M(i\nu_n)\rangle_{imp} = \frac{(4\pi\gamma_e)^2}{V^2}
  \sum_{ \{{\bf k}\} }
   \frac{ ({\bf k}_1 \times {\bf k}'_1)_{+} }{ ({\bf k}_1 - {\bf k}'_1)^2}
   \frac{ ({\bf k}_2 \times {\bf k}'_2)_{-} }{ ({\bf k}_2 - {\bf k}'_2)^2}
   \\
   \times
  \langle {\cal B}(i\nu_n,{\bf k}_1,{\bf k}'_1,{\bf k}_2,{\bf k}'_2) \rangle_{imp},
  \label{K-M-imp-total}
\end{multline}
where $\{{\bf k}\}\equiv {\bf k}_1, {\bf k}'_1, {\bf k}_2, {\bf k}'_2$ and
$\langle ...\rangle_{imp}$ means the average over impurity configurations.
The electronic bubble ${\cal B}(i\nu_n,{\bf k}_1,{\bf k}'_1,{\bf k}_2,{\bf k}'_2)$
is defined in Eq.~(\ref{cccc-averaging}).
For the following discussion it is more convenient to use a different set of wave
vector variables, namely
$ {\bf Q}_a =({\bf k}_a + {\bf k}'_a)/2$ and ${\bf q}_a ={\bf k}_a - {\bf k}'_a$
with $a=1,2$.

The impurity averaging restores the translational invariance of the system and
the result for the electronic bubble, after the spin summation is performed,
can be written as
\begin{eqnarray}
  \langle {\cal B}(i\nu_n,{\bf Q}_1,{\bf q}_1,{\bf Q}_2,{\bf q}_2) \rangle_{imp} &=&
  -2\delta_{{\bf q}_1,-{\bf q}_2} {\cal S}_{{\bf Q}_1,{\bf Q}_2}(i\nu_n,{\bf q}_1).
  \nonumber
\end{eqnarray}
The function ${\cal S}$ has the following general structure\cite{rammer-book}:
\begin{multline}
  {\cal S}_{{\bf Q}_1,{\bf Q}_2}(i\nu_n,{\bf q}) =
  T \sum_{m} G({\bf Q}_{1+}, i\omega_{m+})G({\bf Q}_{1-}, i\omega_m )
  \\
  \times \big[\delta_{{\bf Q}_1,{\bf Q}_2}
  + \frac{1}{V}\Gamma_{{\bf Q}_1,{\bf Q}_2}({\bf q},i\omega_m,i\nu_n)
  \\
   \times G({\bf Q}_{2+}, i\omega_{m+}) G({\bf Q}_{2-}, i\omega_m )\big],
  \label{S-imp-averaged}
\end{multline}
where ${\bf Q }_{a\pm}\equiv {\bf Q}_{a}\pm {\bf q}/2$ for $a=1,2$,
$\omega_{m+}\equiv \omega_m + \nu_n$ and $G({\bf k}, i\omega_m )$ is
the impurity averaged Green's function  defined in Eq.~(\ref{Greens-func-imp}).
The first term in the square brackets in Eq.~(\ref{S-imp-averaged})
corresponds to the bare bubble contribution to $\cal S$.
Here the interaction with impurities is included
through the electronic Green's functions only.
The second term in the square brackets in Eq.~(\ref{S-imp-averaged})
represents the vertex corrections, which are defined in terms of the four point
vertex function $\Gamma_{{\bf Q}_1,{\bf Q}_2}({\bf q},i\omega_m,i\nu_n)$.
This function satisfies the following integral equation\cite{rammer-book}:
\begin{multline}
  \Gamma_{{\bf Q}_1,{\bf Q}_2} = U_{{\bf Q}_1,{\bf Q}_2}+
  \frac{1}{V}\sum_{\bf Q} U_{{\bf Q}_1,{\bf Q}}
  \\
  \times G({\bf Q}_{+}, i\omega_{m+}) G({\bf Q}_{-}, i\omega_m )
  \Gamma_{{\bf Q},{\bf Q}_2},
  \label{eq-for-Gamma}
\end{multline}
where the dependence of $\Gamma$ and $U$ on ${\bf q},i\omega_m$ and $i\nu_n$
is not shown explicitly.

The function $U_{{\bf Q}_1,{\bf Q}_2}({\bf q},i\omega_m,i\nu_n)$,
which appears in Eq.~(\ref{eq-for-Gamma}), contains
all the information about the effective interaction between electrons
arising due to impurity scattering, and can be very complicated in general.
For good metals it is usually a good approximation to consider impurities as
point-like scatterers and take them into account in the Born approximation.
In this case the function
$U_{{\bf Q}_1,{\bf Q}_2}({\bf q},i\omega_m,i\nu_n)$ is essentially a constant,
equal to  $u=n_{imp} V_{imp}^2$. Then Eq.~(\ref{eq-for-Gamma}) has the following
simple solution:
\begin{eqnarray}
\Gamma^{-1}(q,i\omega_m,i\nu_n)&=&
\frac{1}{u}-\frac{1}{V} \sum_{{\bf Q}}
G({\bf Q}_{+}, i\omega_{m+}) G({\bf Q}_{-}, i\omega_m ).
\nonumber
\end{eqnarray}
This result contains the sum of the ladder diagrams and correctly reproduces
the diffusive dynamics of the electrons at low wave vectors and frequencies.

We return now to the correlator of the orbital magnetic fields $K_{+-}^M(i\nu_n)$
in Eq.~(\ref{K-M-imp-total}), and split it into the two pieces:
$K_{+-}^M(i\nu_n) = K_{bare}(i\nu_n) + K_{vert}(i\nu_n)$ corresponding to the
two terms in the square brackets of Eq.~(\ref{S-imp-averaged}). The bare bubble
contribution has the form:
\begin{eqnarray}
 K_{bare}(i\nu_n) &=&
2 \frac{(4\pi\gamma_e)^2}{V^2}\sum_{{\bf q},{\bf Q}}
\left( {\bf q} \times {\bf Q} \right)_{+} \left({\bf q} \times {\bf Q} \right)_{-}
\nonumber
\\
&& \times  \frac{1}{q^4} S({\bf Q},{\bf q},i\nu_n),
\nonumber
\end{eqnarray}
where the function $S$ was defined in the main text in
Eq.~(\ref{el-bubble-im-axis-imp-averaged}), but using the variables
${\bf k}_1 ={\bf Q} + {\bf q}/2$, ${\bf k}_2 ={\bf Q} - {\bf q}/2$.
This bare bubble contribution has been analyzed in Section IIIC.
The vertex correction part can be written as:
\begin{eqnarray}
 K_{vert}(i\nu_n) &=&  2 \frac{(4\pi\gamma_e)^2}{V}
 T \sum_{{\bf q},m} \left( {\bf q} \times {\bf L}({\bf q}) \right)_{+}
\left({\bf q} \times {\bf L}({\bf q})\right)_{-}
\nonumber \\
&& \times  \frac{1}{q^4} \Gamma(q,i\omega_m,i\nu_n),
\nonumber \\
{\bf L}({\bf q}) & \equiv & {\bf L}({\bf q},i\omega_m,i\nu_n)
\nonumber \\
&=& \frac{1}{V} \sum_{{\bf Q}} {\bf Q}
G({\bf Q}_{+}, i\omega_{m+}) G({\bf Q}_{-}, i\omega_m ).
\nonumber
\end{eqnarray}
We see that for an isotropic system ${\bf L}({\bf q}) || {\bf q}$ and
the vertex corrections to the correlator of the orbital magnetic fields
vanish exactly due to its particular vector structure.

\end{document}